\documentclass[a4paper,10pt]{IEEEtran}
\renewcommand{\baselinestretch}{0.99}
\usepackage{pgfplots}
\usetikzlibrary{shapes.multipart,intersections}
\usepackage{cite}
\usepackage{amsmath,amssymb,amsfonts,steinmetz}
\usepackage{algorithmic}
\usepackage{graphicx}
\usepackage{textcomp}
\usepackage{xcolor,flushend}
\def\BibTeX{{\rm B\kern-.05em{\sc i\kern-.025em b}\kern-.08em
    T\kern-.1667em\lower.7ex\hbox{E}\kern-.125emX}}

\usepackage{tikz}
\usetikzlibrary{calc}
\makeatletter
\newcommand{\gettikzxy}[3]{%
  \tikz@scan@one@point\pgfutil@firstofone#1\relax
  \edef#2{\the\pgf@x}%
  \edef#3{\the\pgf@y}%
}
\makeatother

\usepackage[T1]{fontenc}
\usepackage[latin9]{inputenc}
\usepackage{bm}
\usepackage{amsmath}
\usepackage{amssymb}
\usepackage{stmaryrd}
\usepackage{babel}
\usepackage{comment}
\usepackage{graphicx}
\usepackage{algorithm,algorithmic}
\usepackage{xcolor}
\usepackage{gensymb}
\usepackage{cite}
\usepackage{enumitem}
\usepackage{subcaption}
\usepackage{subfig}
\usepackage{url}

\renewcommand{\a}{\mathbf{a}}

\newcommand{\h}{\mathbf{h}}

\newcommand{\n}{\mathbf{n}}

\newcommand{\s}{\mathbf{s}}

\renewcommand{\u}{\mathbf{u}}
\renewcommand{\v}{\mathbf{v}}
\newcommand{\w}{\mathbf{w}}

\newcommand{\y}{\mathbf{y}}
\newcommand{\z}{\mathbf{z}}

\newcommand{\C}{\mathbf{C}}

\renewcommand{\H}{\mathbf{H}}
\newcommand{\I}{\mathbf{I}}

\newcommand{\M}{\mathbf{M}}
\newcommand{\N}{\mathbf{N}}

\newcommand{\R}{\mathbf{R}}

\newcommand{\V}{\mathbf{V}}
\newcommand{\W}{\mathbf{W}}
\newcommand{\X}{\mathbf{X}}
\newcommand{\Y}{\mathbf{Y}}

\newcommand{\Compl}{\mbox{$\mathbb{C}$}}

\renewcommand{\Re}{\mathrm{Re}}

\DeclareMathAlphabet\mathbfcal{OMS}{cmsy}{b}{n}
\usepackage{amsthm}

\begin{document}
\title{Simultaneous Communications and Sensing\\ with Hybrid Reconfigurable Intelligent Surfaces}
\author{\IEEEauthorblockN{
Ioannis Gavras$^1$ and George C. Alexandropoulos$^{1,2}$ 
} 
\\
\IEEEauthorblockA{$^1$Department of Informatics and Telecommunications, National and Kapodistrian University of Athens, Greece}
\IEEEauthorblockA{$^2$Department of Electrical and Computer Engineering, University of Illinois Chicago, USA}
\\
\IEEEauthorblockA{emails: \{giannisgav, alexandg\}@di.uoa.gr}
}

\maketitle
\begin{abstract}
Hybrid Reconfigurable Intelligent Surfaces (HRISs) constitute a new paradigm of truly smart metasurfaces with the additional features of signal reception and processing, which have been primarily considered for channel estimation and self-reconfiguration. In this paper, leveraging the simultaneous tunable reflection and signal absorption functionality of HRIS elements, we present a novel framework for the joint design of transmit beamforming and the HRIS parameters with the goal to maximize downlink communications, while simultaneously illuminating an area of interest for guaranteed localization coverage performance. Our simulation results verify the effectiveness of the proposed scheme and showcase the interplay of the various system parameters on the achievable Integrated Sensing and Communications (ISAC) performance.
\end{abstract}
 
\begin{IEEEkeywords}
Integrated sensing and communications, hybrid reconfigurable intelligent surface, position error bound, THz.
\end{IEEEkeywords}
\pagenumbering{gobble}
\section{Introduction}

The upcoming sixth generation (6G) of wireless networks is expected to offer advanced radar-like sensing capabilities, seamlessly integrated with the evolving use cases of its predecessor~\cite{6G-DISAC_mag}. This new paradigm, known as Integrated Sensing and Communications (ISAC)\cite{mishra2019toward}, has recently garnered significant attention in both research and standardization efforts, with emphasis being given on use cases, channel modeling, and enabling technologies~\cite{RIS_ISAC,18}. To this end, the combination of eXtremely Large (XL) Multiple-Input Multiple-Output (MIMO) systems with THz frequencies shows great promise, offering high angular and range resolution~\cite{THs_loc_survey}. Additionally, the innovative concept of smart radio environments, facilitated by Reconfigurable Intelligent Surfaces (RISs), further advances this vision by enabling programmable control of signal propagation that can be leveraged for various 6G applications, such as ISAC~\cite{RIS_ISAC} and simultaneous localization and radio mapping~\cite{kim2023ris}.

Hybrid RISs (HRISs), equipped with unit elements realizing simultaneous tunable reflection and signal absorption via embedded power splitters as well as limited numbers of Reception (RX) Radio Frequency (RF) chains, were recently proposed in~\cite{hybrid_meta-atom} for dual-functional applications and self-configurability~\cite{alexandropoulos2023hybrid}. For example, the  absorbed portion of impinging signals has been leveraged for channel estimation~\cite{alexandropoulos2020hardware} and localization~\cite{ghazalian2024joint}, balancing hardware cost, power consumption, and computational complexity. HRISs are also emerging as a key technology for ISAC leveraging their large-scale sensing potential due to their large aperture size in terms of the signal wavelength~\cite{alexandropoulos2023hybrid}. Yet, most existing research on RIS-assisted ISAC systems assumes some a priori knowledge of the target locations~\cite{song2023intelligent}. However, in practice, during the target detection phase, this knowledge is hard to acquire using current wireless localization methods~\cite{hua2023secure}. 
%

In this paper, motivated by the HRIS potential for ISAC and the limitations of current sensing approaches with conventional RISs, we focus on the optimization of a novel communications-centric bistatic setup comprising an XL MIMO Base Station (BS) and an HRIS. The objective of the proposed system is to maximize DownLink (DL) communications towards a single User Equipment (UE), while simultaneously illuminating an Area of Interest (AoI)~\cite{RIS_challenges} for a given localization coverage performance. To this end, we derive the Cram\'{e}r-Rao Bound (CRB) for estimating a target's spatial parameters and compute the corresponding Position Error Bound (PEB), which is then used to formulate a PEB constraint on a set of uniformly distributed points within the AoI. Our indcative numerical evaluations over a narrowband subTHz channel model showcase the effectiveness of the proposed ISAC framework, demonstrating our system's dual-functional capabilities as well as the trade-off between communications, localization, and sensing coverage.

\textit{Notations:}
Vectors and matrices are represented by boldface lowercase and uppercase letters, respectively. The transpose, Hermitian transpose, and inverse of $\mathbf{A}$ are denoted as $\mathbf{A}^{\rm T}$, $\mathbf{A}^{\rm H}$, and $\mathbf{A}^{-1}$, respectively. $\mathbf{I}_{n}$, $\mathbf{0}_{n}$, and $\boldsymbol{1}_n$ ($n\geq2$) are the $n\times n$ identity and zeros' matrices and the ones' column vector, respectively. $[\mathbf{A}]_{i,j}$ is $\mathbf{A}$'s $(i,j)$-th element, $\|\mathbf{A}\|$ gives its Euclidean norm, and $\text{Tr}\{\mathbf{A}\}$ its trace. $|a|$ ($\Re\{a\}$) returns the amplitude (real part) of complex scalar $a$. $\mathbb{C}$ is the complex number set, $\jmath\triangleq\sqrt{-1}$ is the imaginary unit, $\mathbb{E}\{\cdot\}$ is the expectation operator, and $\mathbf{x}\sim\mathcal{CN}(\mathbf{a},\mathbf{A})$ indicates a complex Gaussian 
vector with mean $\mathbf{a}$ and covariance matrix $\mathbf{A}$. 

\section{System and Channel Models}
\begin{figure}[!t]
	\begin{center}
	\includegraphics[scale=0.8]{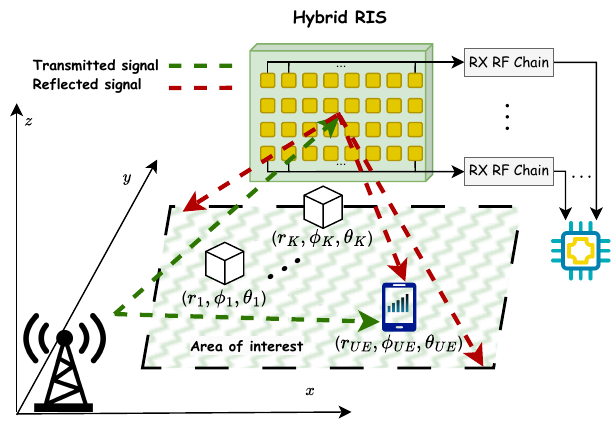}
	\caption{\small{The proposed ISAC system with an HRIS supporting both DL communications and localization coverage across a desired AoI.}}
	\label{fig: system}
	\end{center}
\end{figure}

We consider a wireless system set up consisting of an XL MIMO BS and an HRIS, as illustrated in Fig.~\ref{fig: system}, that wishes to communicate in the DL direction with a single-antenna UE, while simultaneously providing localization coverage over an AoI in the vicinity of the HRIS including $K$ passive radar targets. It is assumed that both the UE and all $K$ targets are located within the near-field region of the HRIS, while this device may or may not be in the near field of the BS. 
The latter node is equipped with a Uniform Planar Array (UPA) consisting of $N\triangleq N_{\rm R}\times N_{\rm E}$ antennas and capable of fully digital Transmit (TX) BeamForming (BF), which positioned in the $xz$-plane at the origin. The HRIS is also modeled as a UPA with $M\triangleq M_{\rm RF}M_{\rm E}$ elements positioned in the $xz$-plane opposite of the BS, with its first element at the point $(r_{\rm RIS},\theta_{\rm RIS},\varphi_{\rm RIS})$ representing respectively the distance from the origin as well as the elevation and azimuth angles. The hybrid meta-atoms at the each column of the HRIS are assumed attached via a dedicated waveguide to a single RX RF chain ($M_{\rm RF}$ in total with $M/M_{\rm RF}\in\mathbb{Z}^+_*$), enabling the absorption of a certain portion of the impinging signals for baseband processing at the HRIS controller~\cite{alexandropoulos2023hybrid}. For the sake of simplicity, we assume that the spacing between adjacent vertical and horizontal elements at both the BS and HRIS UPAs is $\lambda/2$, where $\lambda$ is the communication signal wavelength. We finally adopt the common assumption that the BS and the HRIS are aware of each other's static positions, and that the latter's controller shares its signal observations as well as reflection configurations with the former, via a reliable link~\cite{RIS_challenges}, which is responsible for executing the ISAC design that will be presented later on in Section~III.

The simultaneous reflective and sensing functionality of the HRIS is controlled through $M$ identical power splitters \cite{alexandropoulos2023hybrid,alexandropoulos2020hardware,ghazalian2024joint}, which divide the power of the impinging signal at each hybrid meta-atom in the respective parts. For the sensing operation, to feed the absorbed portion of the impinging signal to the $M_{\rm RF}$ RX RF chains, the HRIS applies the analog combining matrix $\W\in\Compl^{M\times M_{\rm RF}}$, which is modeled as~\cite{ghazalian2024joint}:
\begin{align}
    [\W]_{(l-1)M_{\rm E}+m,j} = \begin{cases}
    w_{l,m},&  l=j\\
    0,              & l\neq j
\end{cases},
\end{align}
where $|w_{l,m}|=1$ (with $m=1,2,\ldots,M$ and $l=1,2,\ldots,M_{\rm RF}$) for each non-zero element in $\W$. In addition, the HRIS reflection configuration is represented by $\boldsymbol{\phi}\in\Compl^{M\times 1}$ with $|[\boldsymbol{\phi}]_m|=1$ $\forall$$m$. Finally, the BS precodes digitally via the channel-dependent BF vector $\v\in\Compl^{N\times 1}$ each complex-valued symbol $s$ such that $\mathbb{E}\{\|\v s\|^2\}\leq P_{\rm max}$, where $P_{\rm max}$ represents the maximum transmission power.

\subsection{Near-Field Channel Model}
We consider wireless operations in the sub-THz frequency band, according to which the $M\times N$ complex-valued channel matrix between the BS and the HRIS is modeled as~\cite{THs_loc_survey}:
\begin{align}\label{eq: BR}
    [\H_{\rm BR}]_{(l-1)M_{\rm E}+m,n} \triangleq \alpha_{l,m,n} \exp\left(\frac{\jmath2\pi}{\lambda} r_{l,m,n}\right)
\end{align}
with $r_{l,m,n}$ denoting the distance between each $m$-th element of each $l$-th RX RF chain at the HRIS and the $n$-th (with $n=1,2,\ldots,N$) BS antenna. 
In addition, $\alpha_{l,m,n}$ represents the respective attenuation factor including the molecular absorption coefficient $\kappa_{\rm abs}$, which is defined as:
\begin{align}\label{eq: atn}
    \alpha_{l,m,n} \triangleq \frac{\lambda}{4\pi r_{l,m,n}}\sqrt{F(\theta_{l,m,n})}\exp\left(-\frac{\kappa_{\rm abs}r_{l,m,n}}{2}\right)
\end{align}
with $F(\cdot)$ being this meta-atom's radiation profile \cite{FD_HMIMO_2023} that depends on the respective elevation angle $\theta_{l,m,n}$. It is noted that the geometric relationships between $r_{l,m,n}$ and $\theta_{l,m,n}$ $\forall l,m,n$ can be computed as described in \cite{FD_HMIMO_2023}. 

Let the spherical coordinates of the $K$ potentially present passive radar targets within the AoI, as shown in Fig.~\ref{fig: system}, be denoted by $\{(r_1,\theta_1,\varphi_1),(r_2,\theta_2,\varphi_2),\ldots,(r_K,\theta_K,\varphi_K)\}$. The end-to-end channel matrix including the impinging/reflected components to/from the $K$ targets and the UE, when considered as point sources, can be expressed as follows: 
\begin{align}\label{eq:H_R}
    \H_{\rm R} \triangleq& \sum\limits_{k=1}^{K}\beta_k \a_{\rm RX}(r_k,\theta_k,\varphi_k)\a_{\rm TX}^{\rm H}(r_k,\theta_k,\varphi_k)\\
    &+\beta_{\rm UE} \a_{\rm RX}(r_{\rm UE},\theta_{\rm UE},\varphi_{\rm UE})\a_{\rm TX}^{\rm H}(r_{\rm UE},\theta_{\rm UE},\varphi_{\rm UE}),\nonumber
\end{align}
where $\beta_k$ and $\beta_{\rm UE}$ represent the complex-valued reflection coefficients for each $k$-th (with $k=1,2,\ldots,K$) radar target and the UE, respectively, and $\forall$$l,m,n$:
\begin{equation}\label{eq: response_vec}
\begin{split}
    &[\a_{\rm TX}(r,\theta,\varphi)]_n \triangleq a_n\exp\Big({\jmath\frac{2\pi}{\lambda}r_n}\Big),\\
    &[\a_{\rm RX}(r,\theta,\varphi)]_{(l-1)M_{\rm E}+m} \triangleq a_{l,m}\exp\Big({\jmath\frac{2\pi}{\lambda}r_{l,m}}\Big),
\end{split}
\end{equation}
where $a_n$ and $r_n$ denote the attenuation factor and the distance between each $n$-th BS antenna element and either a passive target or the UE. Similarly, $a_{l,m}$ and $r_{l,m}$ are the attenuation factor and distance between each $m$-th element of each $l$-th RX RF chain of the HRIS and either a target or the UE. All latter parameters are computed as in \eqref{eq: atn} and~\cite{FD_HMIMO_2023} according to each $k$-th passive target's or UE's position. Finally, the channel between the BS and the UE is represented as $\h_{\rm BU}\triangleq\a_{\rm TX}(r_{\rm UE},\theta_{\rm UE},\varphi_{\rm UE})$, while the channel between the UE and the HRIS as $\h_{\rm RU}\triangleq\a_{\rm RX}^{\rm T}(r_{\rm UE},\theta_{\rm UE},\varphi_{\rm UE})$ with $(r_{\rm UE},\theta_{\rm UE},\varphi_{\rm UE})$ denoting the spherical coordinates of the UE.



\subsection{Received Signal Models}
The baseband received signal at the UE, consisting of the signal travelling through the direct link and that through the HRIS, can be mathematically expressed as follows:
\begin{align}\label{eq:UE_received_signal}
    y\triangleq\left(\h_{\rm BU}+(1-\rho)\h_{\rm RU}{\rm diag}(\boldsymbol{\phi})\H_{\rm BR}\right)\v s+n,
\end{align}
where $n\sim\mathcal{CN}(0,\sigma^2)$ is the Additive White Gaussian Noise (AWGN) and $(1-\rho)$ represents the absorption portion of the impinging signal at every hybrid meta-atom, with $\rho\in[0,1]$ being the common power splitting ratio. 

By assuming $T$ BS transmissions per coherent channel block, and that the static BS-HRIS channel can be completely cancelled, the baseband received signal $\Y\in\Compl^{M_{\rm RF}\times T}$ at the outputs of the $M_{\rm RF}$ RX RF chains of the HRIS can be mathematical expressed as follows:
\begin{align}\label{eq: y_ref}
    \Y = [\y(1),\y(2),\ldots,\y(T)] \triangleq \rho\W^{\rm H}\H_{\rm R}\v\s+\N,
\end{align}
where $\s \triangleq [s(1),s(2),\ldots,s(T)]\in\Compl^{1\times T}$ and $\N \triangleq [\n(1),\n(2),\ldots,\n(T)]\in\Compl^{M_{\rm RF}\times T}$, with $\n(t)\sim\mathcal{CN}(0,\sigma^2\I_{M_{\rm RF}})$ $\forall$$t=1,2,\dots,T$, being the vector with the transmitted symbols and the matrix with the AWGN vectors per received symbol, respectively. 

\section{Proposed HRIS-Enabled ISAC Framework}
In this section, we first derive the PEB with respect to the HRIS reception capability, and then present the design objective for the proposed HRIS-enabled simultaneous communications and sensing system together with our solution for the TX BF and HRIS parameters.

\subsection{PEB Analysis}
It is evident from \eqref{eq: y_ref}'s inspection that, for a coherent channel block involving $T$ unit-powered symbol transmissions, yields $T^{-1}\s\s^{\rm H}=1$ and the received signal at the output of the HRIS's RX RF chains is distributed as $\y\sim\mathcal{CN}(\boldsymbol{\M},\R_n)$, with mean $\boldsymbol{\M} \triangleq \rho\W^{\rm H}\H_{\rm R}\v\s$ and covariance matrix $\R_n\triangleq\sigma^2\I_{M_{\rm RF}}$. In the context of estimating the $k$-th passive radar target positioned at $\boldsymbol{\zeta} \triangleq [r,\theta,\varphi]^{\rm T}$, each $(i,j)$-th element (with $i,j=1,2,$ and $3$) of its $3\times3$ Fisher Information Matrix (FIM), $\mathbfcal{I}$, can be calculated as follows~\cite{kay1993fundamentals}:
\begin{align}
    \nonumber[\mathbfcal{I}]_{i,j} \!\triangleq\! 2\Re\left\{\!\frac{\partial \boldsymbol{\M}^{\rm H}}{\partial[\boldsymbol{\zeta}]_i}\R_n^{-1}\frac{\partial \boldsymbol{\M}}{\partial[\boldsymbol{\zeta}]_j}\!\right\}\!+\!\text{Tr}\left\{\!\R_n^{-1}\frac{\partial\R_n}{\partial[\boldsymbol{\zeta}]_i}\R_n^{-1}\frac{\partial\R_n}{\partial[\boldsymbol{\zeta}]_j}\!\right\}\!.
\end{align}
Since $\R_n$ is independent of $\boldsymbol{\zeta}$, it holds that $\nabla_{\boldsymbol{\zeta}}\R_n = \mathbf{0}_{3\times1}$. Therefore, each element of the FIM depends solely on the mean, and $\forall i$ the following result can be easily deduced: $\frac{\partial\boldsymbol{\M}}{\partial[\boldsymbol{\zeta}]_i} = \rho\W^{\rm H}\frac{\partial\H_{\rm R}}{\partial[\boldsymbol{\zeta}]_i}\v\s$.
Consequently, each diagonal element of the FIM can be re-written as follows:
\begin{align}
    [\mathbfcal{I}]_{i,i} = \frac{2\rho^2}{\sigma^2}\Re\Bigg\{\s^{\rm H}\v^{\rm H}\frac{\partial\H_{\rm R}^{\rm H}}{\partial[\boldsymbol{\zeta}]_i}\W\W^{\rm H}\frac{\partial\H_{\rm R}}{\partial[\boldsymbol{\zeta}]_i}\v\s\Bigg\}.\nonumber
\end{align}
Combining all above together, the PEB for each $k$-th target with true, but unknown, position $\boldsymbol{\zeta}$ can be expressed as:
\begin{align}\label{eq: PEB}
\text{PEB}_{\boldsymbol{\zeta}} \triangleq \sqrt{{\rm CRB}_{\boldsymbol{\zeta}}}=
\sqrt{{\rm Tr}\left\{\mathbfcal{I}^{-1}\right\}}. 
\end{align}

\subsection{Problem Formulation and Solution}
Our goal is to jointly optimize the reconfigurable parameters of the BS and the HRIS to maximize the achievable DL rate towards the single UE, while ensuring a localization coverage performance threshold with respect to an AoI in the near-field region of the HRIS. For the latter requirement, we particularly focus on our system's ability to maintain a consistent level of target estimation accuracy across the entire AoI. To this end, we discretize the AoI into a finite set of points, each represented by the polar coordinates $\boldsymbol{\zeta}_q\triangleq (r_q, \varphi_q, \theta_q)$ with $q=1,2,\ldots,\mathcal{Q}$. In mathematical terms, our HRIS-enabled ISAC design objective is expressed as follows:
\begin{align}
        \mathcal{OP}\!:& \nonumber\underset{\substack{\W,\boldsymbol{\phi},\v}}{\max} \,\,\log_2\left(1+\sigma^{-2}\|\widehat{\h}_{\rm DL}\v\|^2\right)\\
        &\nonumber\text{s.t.}\, {\rm PEB}_{q}\leq\gamma_s\,\forall q, |[\w]_{l,m}|=1 \,\forall l,m, \left\|\v\right\|^2\leq P_{\rm max},
\end{align} 
where we have assumed unit power symbol transmissions and $\gamma_s$ represents the desired localization accuracy. In addition, $\widehat{\h}_{\rm DL}\triangleq\widehat{\h}_{\rm BU}+(1-\rho)\widehat{\h}_{\rm RU}{\rm diag}(\boldsymbol{\gamma})\widehat{\H}_{\rm BR}$ denotes the estimation of the DL channel towards the UE via the model in~\eqref{eq:UE_received_signal} using the estimations $\widehat{\h}_{\rm BU}$ and $\widehat{\h}_{\rm RU}$ for $\h_{\rm BU}$ and $\h_{\rm RU}$, respectively, which can be obtained via typical channel estimation schemes~\cite{alexandropoulos2020hardware}. Clearly, $\mathcal{OP}$ is a non-convex and highly coupled problem, thus, finding its optimal solution is extremely difficult. 
To this end, following the approach in~\cite{spawc2024}, we leverage the positive semidefinite nature of the FIM and the lower bound ${\rm Tr}\{\mathbfcal{I}^{-1}\}\geq\frac{9T^2}{{\rm Tr}\{\mathbfcal{I}\}}$, which results from the previously derived PEB expression in~\eqref{eq: PEB}, to reformulate each PEB constraint as follows:  
\begin{align}\label{eq: PEB_simp}
 \sum_{i=1}^{3}\Re\Bigg\{\v^{\rm H}\frac{\partial\H_{\rm R}^{\rm H}}{\partial[\boldsymbol{\zeta}_q]_i}\W\W^{\rm H}\frac{\partial\H_{\rm R}}{\partial[\boldsymbol{\zeta}_q]_i}\v\Bigg\}\geq\frac{\sigma^2}{2\gamma_s^2\rho^2T}.   
\end{align}
However, it is evident from this expression that certain spatial dimensions may contribute more significantly to the PEB than others. In fact, for our near-field scenario, it is easier to estimate the azimuth and elevation angles of a target rather than its range, i.e., the beamdepth exceeds the beamwidth~\cite{THs_loc_survey}. 
Hence, to achieve consistent estimation accuracy across all dimensions, we further decompose \eqref{eq: PEB_simp} into its individual components. We particularly reformulate $\mathcal{OP}$ as follows:
\begin{align}
        \nonumber\mathcal{OP}_1\!:&\underset{\substack{\substack{\W,\boldsymbol{\phi},\v,\\\{t_{q,i}\}_{q,i=1}^{\mathcal{Q},3}}}}{\max} \|\widehat{\h}_{\rm DL}\v\|^2+\sum_{q=1}^\mathcal{Q}\sum_{i=1}^3t_{q,i}\\
        &\nonumber\text{\text{s}.\text{t}.}\,\nonumber\Re\Bigg\{\v^{\rm H}\frac{\partial\H_{\rm R}^{\rm H}}{\partial[\boldsymbol{\zeta}_q]_i}\W\W^{\rm H}\frac{\partial\H_{\rm R}}{\partial[\boldsymbol{\zeta}_q]_i}\v\Bigg\}\geq \frac{t_{q,i}\sigma^2}{2\rho^2T},\,\,\\&\nonumber\quad\,\,\, t_{q,i}\leq\gamma_s^{-2}\,\,\forall q,i,\, |[\w]_{l,m}|=1\forall l,m,\, \|\v\|^2\leq P_{\rm max},
\end{align} 
where we have used the set of auxiliary variables $t_{q,i}$ $\forall q,i$, to ensure that, if a PEB constraint cannot be fully satisfied, the solution maximizes compliance as much as possible. 
\begin{figure*}[!t]
  \begin{subfigure}[t]{0.4\textwidth}
  \centering
    \includegraphics[width=\textwidth]{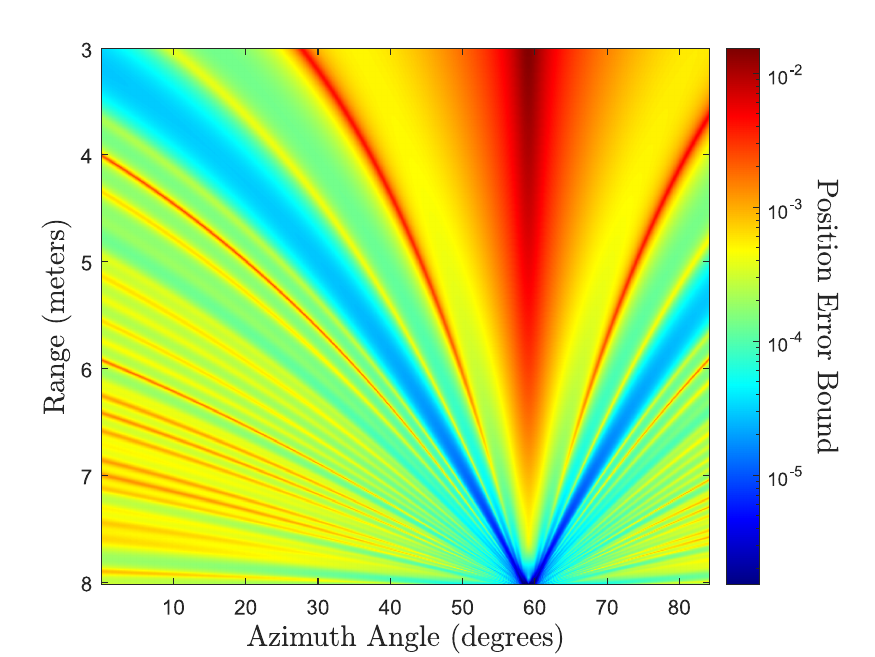}
    \caption{$\rho = 0.2$ and $\mathcal{Q}=5$.}
    \label{fig:PEB}
  \end{subfigure}\hfill
  \begin{subfigure}[t]{0.4\textwidth}
  \centering
    \includegraphics[width=\textwidth]{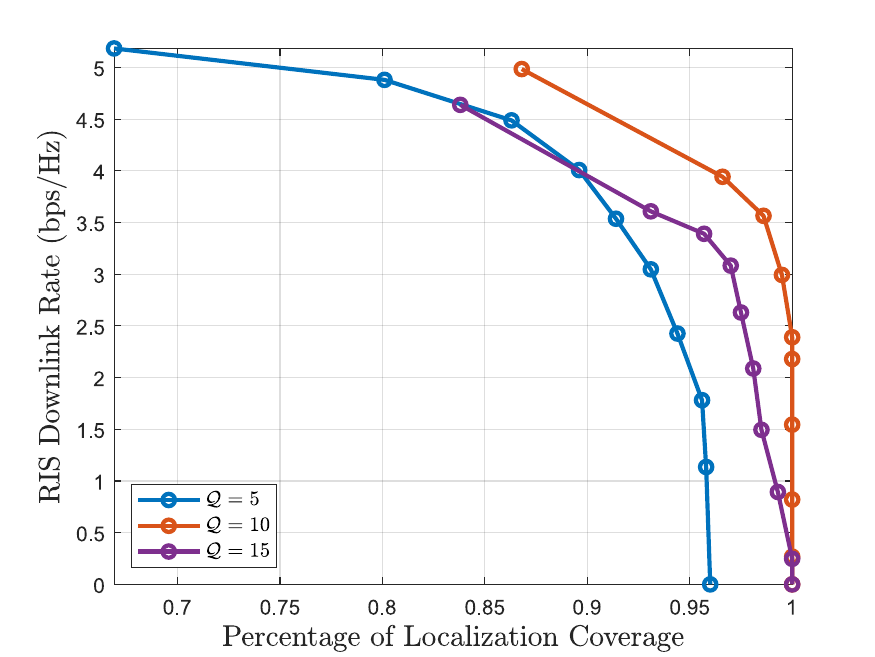}
    \caption{$\rho\in(0,1]$ and $\mathcal{Q}=\{5,10,15\}$.}
    \label{fig:Tradeoff}
  \end{subfigure}
  \caption{\small{PEB (left) and the trade-off between achievable DL rate and localization coverage performance (right) for the proposed ISAC system with $P_{\rm \max}=16$ dBm and $\gamma_s=10^{-3}$ including an HRIS with $M_{\rm RF} = 4$ RX RF chains each connected to $M_{\rm E} = 64$ hybrid meta-atoms.}\vspace{-0.4cm}}
  \label{fig:FIG1}
\end{figure*}

We now adopt alternating optimization to solve $\mathcal{OP}_1$. In particular, we can effectively design the BS's digital BF vector and the combining matrix of the HRIS via performing semi-definite relaxation in the following optimization problems:
\begin{align}
        \mathcal{OP_{\rm \V}}:&\nonumber\underset{\substack{\V,\{t_{q,i}\}_{q,i=1}^{\mathcal{Q},3}}}{\max} \|\V\widehat{\h}_{\rm DL}^{\rm H}\widehat{\h}_{\rm DL}\|^2+\sum_{q=1}^\mathcal{Q}\sum_{i=1}^3t_{q,i}\\
        &\nonumber\text{\text{s}.\text{t}.}\,\nonumber{\rm Tr}\left\{\Re\Bigg\{\V\frac{\partial\H_{\rm R}^{\rm H}}{\partial[\boldsymbol{\zeta}_q]_i}\W\W^{\rm H}\frac{\partial\H_{\rm R}}{\partial[\boldsymbol{\zeta}_q]_i}\Bigg\}\right\}\geq \frac{t_{q,i}\sigma^2}{2\rho^2T},
        \\&\nonumber\quad\,\,\, t_{q,i}\leq\gamma_s^{-2}\,\,\forall q,i, {\rm Tr}\{\V\}\leq P_{\rm max}, \V\succeq0.
\end{align}
\begin{align}
        &\mathcal{OP}_{\X_l}:\nonumber\underset{\substack{\X_l,\{t_{q,i}\}_{q,i=1}^{\mathcal{Q},3}}}{\max} \quad\sum_{q=1}^\mathcal{Q}\sum_{i=1}^3t_{q,i}\\
        &\nonumber\text{\text{s}.\text{t}.}\,\nonumber{\rm Tr}\left\{\Re\Bigg\{\X_l\left[\frac{\partial\H_{\rm R}}{\partial[\boldsymbol{\zeta}_q]_i}\v\v^{\rm H}\frac{\partial\H_{\rm R}^{\rm H}}{\partial[\boldsymbol{\zeta}_q]_i}\right]_{k_l,k_l}\Bigg\}\right\}\geq \frac{t_{q,i}\sigma^2}{2\rho^2T},\,\,\\&\nonumber\quad\,\,\, t_{q,i}\leq\gamma_s^{-2}\,\,\forall q,i,\,{\rm diag}(\X_l) = \boldsymbol{1}_{M},\,\X_l\succeq0.
\end{align} 
In the former problem, we have set $\V\triangleq\v\v^{\rm H}$ to relax the rank-one constraint on the TX BF, and consequently, reformulated this parameter's optimization as a convex problem, which can be efficiently solved using available solvers. Clearly, due to $\mathcal{OP}{\rm \v}$'s structure, a rank-one solution is not feasible~\cite{huang2009rank}. Hence, we apply a low-rank approximation according to which the BF vector $\v$ is recovered as $\v = \u_1\sqrt{\sigma_1}$, where $\sigma_1$ is the largest singular value of $\V$ and $\u_1$ its corresponding singular vector. In the latter problem for each $l$-th RX RF chain at the HRIS,
we have defined $\X_l\triangleq\W(k_l:lM_{\rm E},l)\W^{\rm H}(k_l:lM_{\rm E},l)$ with $k_l\triangleq(l-1)M_{\rm E}+1$ and removed the rank-one constraint to obtain a convex formulation, which can be solved as $\W(k_l:lM_{\rm E},l)=\exp\left(\jmath\angle\z_1\right)$ with $\z_1$ being the principal singular vector of $\X_l$. The HRIS reflection configuration vector can be finally designed via the following optimization problem that uses the definition $\boldsymbol{\phi}=e^{\jmath\boldsymbol{\varphi}}$ with the vector $\boldsymbol{\varphi}$ including the $M$ tunable reflection coefficients:
\begin{align}
        \mathcal{OP_{\rm \boldsymbol{\phi}}}:
        &\,\nonumber\underset{\substack{\boldsymbol{\phi}}}{\max} \quad\|(\widehat{\h}_{\rm BU}+(1-\rho)\widehat{\h}_{\rm RU}{\rm diag}(\boldsymbol{\phi})\widehat{\H}_{\rm BR})\v\|^2\\
        &\nonumber\text{\text{s}.\text{t}.}\,\, -\frac{\pi}{2}\leq[\boldsymbol{\varphi}]_m\leq\frac{\pi}{2}\,\forall m=1,2,\ldots,M,
\end{align}
This problem can be efficiently solved with any of the available optimization approaches for conventional RISs~\cite{Tsinghua_RIS_Tutorial}.

Putting all above together, $\mathcal{OP}$ is iteratively solved by sequentially computing $\v$, $\W$, and $\boldsymbol{\phi}$ as the solutions of $\mathcal{OP}_{\V},\mathcal{OP}_{\X_l}$$\forall l$, and $\mathcal{OP}_{\boldsymbol{\phi}}$, respectively, until a convergence criterion is satisfied or a maximum iteration number is reached. 

\section{Numerical Results and Discussion}\label{sec: num}

In this section, we numerically evaluate the simultaneous communications and localization coverage performance of the proposed HRIS-enabled ISAC framework. We have simulated a narrowband subTHz setup with bandwidth $B = 150$ kHz centered at the frequency of $120$ GHz, where coherent channel blocks span $T=200$ transmissions. For the estimation of the passive radar targets within the near-field AoI of the HRIS, we have used the modified multiple signal classification algorithm presented in~\cite{FD_HMIMO_2023}. The AoI was defined at the fixed elevation angle of $\theta = 30^\circ$, azimuth angle of $\varphi \in [20^\circ, 80^\circ]$, and range of $r\in[0.62\sqrt{D^3/\lambda},2D^2/\lambda]$ meters with $D$ being the HRIS diagonal length (i.e., at HRIS's Fresnel region). Both the UE and the $K=2$ radar targets were randomly positioned within the AoI, while the HRIS was considered placed at the point with $\theta_{\rm RIS} = 30^\circ$, $\varphi_{\rm RIS} = 60^\circ$, and $r_{\rm RIS} = 8$ meters. We have used $500$ Monte Carlo trials for all our simulation results, considering a BS equipped with a $2 \times 8$ BS UPA and an HRIS with $M_{\rm RF} = 4$ RX RF chains each connected to $M_{\rm E} = 64$ hybrid meta-atoms. We have set AWGN's variance as $\sigma^2 = -174 + 10\log_{10}(B)$ and the channel coefficients $\beta_{\rm UE}$ and $\beta_k$ $\forall$$k$ were randomly selected with unit amplitude. For the evaluation of the localization coverage, we have computed the coverage capability as the percentage of discrete points in the AoI ($1000$ equidistant points) that satisfy the PEB constraint.

Figure~\ref{fig:FIG1} illustrates the average PEB performance, achievable DL rate, and localization coverage performance for the proposed HRIS-enabled ISAC system design operating with $P_{\rm \max}=16$ dBm and $\gamma_s=10^{-3}$. In Fig.~\ref{fig:PEB}, considering $\rho = 0.2$ and a near-field AoI with respect to the HRIS consisting of $\mathcal{Q}=5$ equidistant discrete points, it is demonstrated that the desired PEB threshold is met throughout approximately the $80\%$ of the AoI. The remaining regions of this area, where the PEB threshold is not met, are primarily located directly in front of the HRIS, and depend on the Signal-to-Noise Ratio (SNR) value as well as the considered low absorption level $\rho$ per meta-atom.  Recall that the HRIS is positioned at the angle $\theta_{\rm RIS} = 30^{\circ}$ and is tasked with providing coverage over the same elevation plane. This results in range ambiguity in the azimuth/elevation plane directly in front of the HRIS, where this bistatic setup struggles to differentiate target ranges along the boresight of the HRIS leading to PEB performance degradation. 
The trade-off between the achievable DL rate and the localization coverage performance is depicted in Fig.~\ref{fig:Tradeoff} for $10$ different absorption levels $\rho$ equispaced in $(0,1]$ and different numbers of discretization points $\mathcal{Q}=\{5,10,15\}$ of the AoI. As expected, it is shown that, by increasing the absorption level, the localization coverage improves, while the communication performance is degraded. It is also demonstrated that better localization coverage can be achieved even with a low absorption level if the discretization level of the AoI increases. However, this happens up to a certain $\mathcal{Q}$ value that, in this setup, seems to be close enough to $N-1$ with $N$ being the total number of BS antennas; recall that, in this figure, we have set $N=16$ and the number of HRIS elements is $M=256$.  
\begin{figure}[!t]
	\begin{center}
	\includegraphics[scale=0.5]{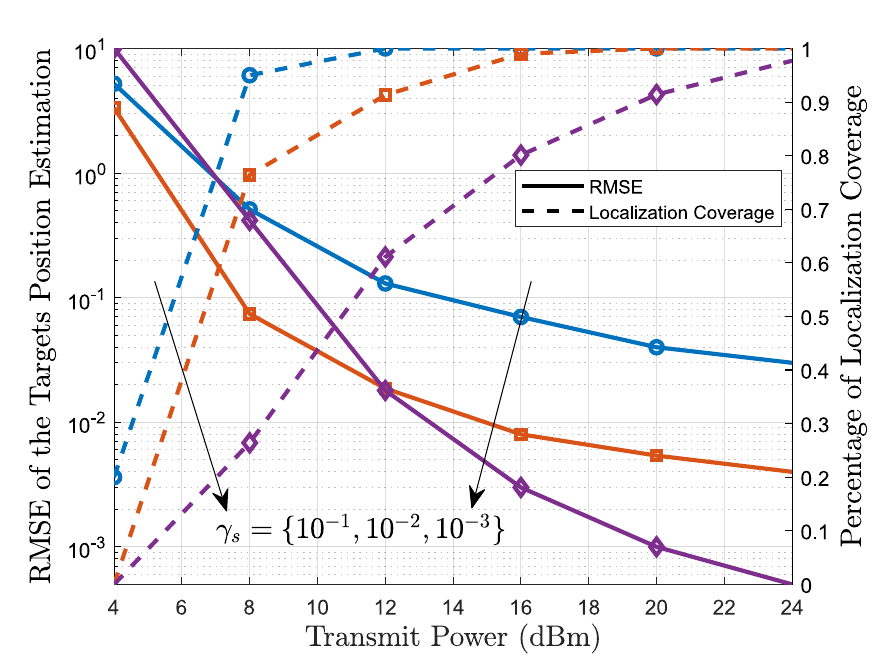}
	\caption{\small{RMSE of target estimation and localization coverage for $\rho = 0.2$, $\mathcal{Q} = 5$, and different localization accuracy thresholds $\gamma_s$.}}
	\label{fig:RMSE}
	\end{center}
\end{figure}

Figure~\ref{fig:RMSE} depicts the Root Mean Squared Error (RMSE) of the estimation of $K=2$ targets lying within the AoI as well as the localization coverage performance for $\rho = 0.2$, $\mathcal{Q}=5$, and different $P_{\rm \max}$ values. It can be seen that, as the SNR increases, both investigated metrics improve. As expected, for increased localization coverage, the estimation performance per position gets degraded, which also leads to decreased achievable DL rate from Fig.~\ref{fig:Tradeoff}.
It is also shown that, for stricter $\gamma_s$ thresholds and lower SNR values, both target estimation and wide localization coverage become extremely challenging. On the other hand, as the SNR increases, the rate of target estimation improvement accelerates, leading to improved RMSE performance together with wider desired localization coverage.

\section{Conclusion}
This paper introduced a communications-centric bistatic setup comprising an XL MIMO BS and an HRIS, which was designed to maximize DL communications simultaneously with ensuring guaranteed localization coverage performance across a desired AoI in the near-field of the HRIS. We derived the PEB for the positions of passive targets within the AoI and modeled this area within our ISAC problem formulation as series of PEB constraints for reliable localization coverage. Our results validated the proposed approach and highlighted trade-offs between communications, sensing, and coverage.

\section*{Acknowledgments}
This work has been supported by the SNS JU projects TERRAMETA and 6G-DISAC under the EU's Horizon Europe research and innovation programme under Grant Agreement numbers 101097101 and 101139130, respectively. TERRAMETA also includes top-up funding by UKRI under the UK government's Horizon Europe funding guarantee.

\bibliographystyle{IEEEtran}
\bibliography{ms}

\end{document}